# SOK: Blockchain for Provenance

Asma Jodeiri Akbarfam
ajodeiriakbarfam@augusta.edu
School of Computer and Cyber Sciences, Augusta University
Augusta, GA, USA

Hoda Maleki
hmaleki@augusta.edu
School of Computer and Cyber Sciences, Augusta University
Augusta, GA, USA

## Abstract

Provenance, which traces data from its creation to manipulation, is crucial for ensuring data integrity, reliability, and trustworthiness. It is valuable for single-user applications, collaboration within organizations, and across organizations. Blockchain technology has become a popular choice for implementing provenance due to its distributed, transparent, and immutable nature. Numerous studies on blockchain designs are specifically dedicated to provenance, and specialize in this area. Our goal is to provide a new perspective in blockchain based provenance field by identifying the challenges faced and suggesting future research directions. In this paper, we categorize the problem statement into three main research questions to investigate key issues comprehensively and propose a new outlook on the use of blockchains. The first focuses on challenges in non-collaborative, single-source environments, the second examines implications in collaborative environments and different domains such as supply chain, scientific collaboration and digital forensic, and the last one analyzes communication and data exchange challenges between organizations using different blockchains. The interconnected nature of these research questions ensures a thorough exploration of provenance requirements, leading to more effective and secure systems. After analyzing the requirements of provenance in different environments, we provide future design considerations for provenance-based blockchains, including blockchain type, query mechanisms, provenance capture methods, and domain-specific considerations. We also discuss future work and possible extensions in this field.

## 1 Introduction

Data provenance involves tracing and authenticating the origin, custody, and history of data artifacts and plays a critical role in ensuring collaboration integrity by addressing tampering and data manipulation. Existing provenance systems operate under the assumption that data collection and storage mechanisms are secure and often depend on trusted third parties for verification. These assumptions require robust guarantees that both the mechanisms and the environment remain uncompromised [61]. In practice, this is often impractical, as achieving complete security protection is rarely possible. Consequently, blockchain technology has been explored for recording data provenance due to its distributed nature, integrity protection, and tamper-evident properties [65]. The use of blockchain for provenance varies. Some designs cater to clients tracking their data, raising issues such as online or offline querying and determining who can query and verify the provenance. Other designs support collaborative efforts in domains where collaboration is essential, such as general data protection regulation (GDPR), [58], IoT, supply chain management [23], machine learning [51], cloud computing [89], scientific workflows, legal scenarios, and digital forensics [36, 68, 77]. In these designs it is crucial to address provenance, collaboration, privacy, confidentiality, and domain-specific issues such as legitimate product registration in supply chains, cross-border jurisdictions in digital forensics, and workflow complexity in scientific research. Organizations with distinct blockchains seeking to collaborate face additional challenges. They often encounter issues due to the independent operation of private or public blockchains, effectively isolating them as separate entities [60]. This isolation highlights the importance of chain interoperability, a concept not initially considered in the early stages of blockchain technology.

The idea of cross-chain interoperability, introduced in 2014 by the Tendermint team, brought the concept of communication and interaction between relatively independent blockchain ledgers to researchers [43, 72]. Cross-chain interoperability involves achieving communication and interaction between independent blockchain ledgers, covering assets and data [60]. However, this integration introduces three key challenges. Firstly, achieving interoperability without a trusted third party is considered unattainable, due to the need for secure and reliable communication, requiring either centralized or decentralized trust mechanisms [18, 44]. Secondly, structural differences in cross-chain processes designed by various solutions pose standardization challenges, necessitating a unified approach to enhance functionality and scalability [82]. Lastly, managing cross-chain historical data, particularly provenance, presents difficulties in overcoming data isolation for accessibility in blockchain activities and analyses [82].

Consequently, leveraging blockchain for provenance requires addressing provenance in single-entity environments, where users seek to guarantee data integrity, as well as in organizations with collaborative multi-user blockchain environments or across multiple organizations collaborating in a multi-chain setting. To explore these aspects, this paper categorizes the problem into three main research questions (RQs):

- **RQ1**: What are the primary challenges faced in implementing blockchain-based data provenance frameworks in non-collaborative, single-source environments?
- **RQ2**: What are the implications of utilizing blockchain technology to ensure data provenance in collaborative environments?
- **RQ3**: How can we overcome challenges to facilitate secure and efficient communication and data exchange while capturing provenance between organizations with different blockchains?

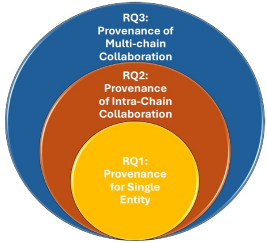

Figure 1: Interrelation of Research Questions

In RQ1, we explore cloud environments vulnerability to accidental corruption or intentional forgery, highlighting the importance of provenance for individual entities where their files are stored without collaboration. We examine the literature and needs of such systems. Addressing RQ1 is essential for setting a solid foundation for understanding provenance needs. Once the baseline provenance needs for individual entities are understood, the next step is to explore how these needs evolve when multiple entities collaborate . Thereby with RQ2, we aim to recognize the different domains including scientific collaboration, supply chain management, healthcare systems, machine learning, and digital forensics, understand their goals and designs, and identify key factors to set an environment for effective collaboration and capture provenance accurately. The goal of RQ3 is to analyze the communication needs for blockchains to collaborate and the provenance requirements of these systems. As depicted in Figure 1, these RQs are sequential and interconnected, indicating that addressing a preceding RQ is necessary to tackle the subsequent one. After reviewing the literature and contributions to these RQs, we aim to provide future design considerations such as blockchain choice, domain-specific design, provenance capture and query, evaluation and selection suggestions. Additionally, we offer a perspective for future work in the field of blockchain-based provenance, highlighting the less explored topics and subtopics related to this area which are crucial for practical, secure and efficient solutions. Our broader analysis of blockchain-based provenance systems differs from earlier surveys and studies that focus on specific technical and security aspects within a narrower scope such as consensus for cloud provenance [76].

This paper is structured as follows: Section 2 provides an overview of essential background concepts. Sections 3, 4 and 5 delve deeper into the problem statement, exploring the RQs and related challenges. Section 6 explores the future design and research directions for blockchain-based provenance. Section 7 concludes the findings.

## 2 Background

In this section, we delve into the background information.

### 2.1 Blockchain

A blockchain is a chain of blocks as depicted in Figure 2 which functions as a decentralized and distributed ledger system that securely records transactions and stores information across multiple network nodes [4–6]. Trust is established through consensus mechanisms such as Proof of Work (PoW) [55], or Proof of Stake (PoS) [41] or Byzantine Fault Tolerant (BFT) algorithms[19] , which ensure that all transactions are verified and agreed upon by the majority of nodes. Immutability, a key feature of blockchain, ensures data integrity and tamper resistance. This property relies on the storage of the Merkle root and the hash of the previous block, forming a chain of linked blocks where any alteration to a previous block invalidates subsequent ones [17].

Blockchain technology encompasses various types, including public and private blockchains. Public blockchains, such as Bitcoin and Ethereum, are open to all participants, while private blockchains restrict access to specific groups, commonly used in enterprise settings for enhanced privacy and control [7, 15]. Additionally, some blockchains incorporate smart contracts [20, 73], self-executing programs stored on the blockchain, enabling decentralized decision-making and automated processes according to predefined conditions [79]. These features collectively contribute to the effectiveness and widespread adoption of blockchain technology across different industries.

### 2.2 Provenance

Data provenance, also known as lineage, is the documentation of data origins and its journey throughout its life cycle. It includes metadata that describes the origins, history, and evolution of an end product. Provenance encompasses various entities, such as data, processes, activities, and users, involved in the entire data life cycle. Provenance in different domains such as supply chain, digital forensics, and scientific collaboration [12, 36, 57] can have different meanings, as illustrated in Table 1. With the exponential growth of digital data being generated, copied, transferred, and manipulated online, provenance has become increasingly vital in ensuring security. Various methods exist for recording provenance, as illustrated in Figure 3. One scenario involves users having direct access to the data store and sending the metadata to the provenance storage. Another scenarios allows users direct access to the data, with the data store itself sending the metadata to the provenance storage. Alternatively, when users do not have direct access, a third party, whether centralized or decentralized, authenticates their access to the data [9] and sends the metadata to the provenance storage. Additionally in some cases, provenance information can be stored in different locations or sent by multiple users.

Table 1: Provenance Record Fields

| Product Supply Chain | Digital Forensics | Scientific Collaboration |
| --- | --- | --- |
| Unique Product ID | Case Number | Task ID |
| Batch or Lot Number | Investigation Stage | Workflow ID |
| Manufacturing and Expiration Date | Case Start Date | Execution Time |
| Travel Trace | Case Closure Date | User ID |
| Product Type or Category | File Types | Input Data |
| Manufacturer ID | Access Patterns | Output Data |
| Quick Access URL or QR Code | Files Dependency | Invalidated Results |

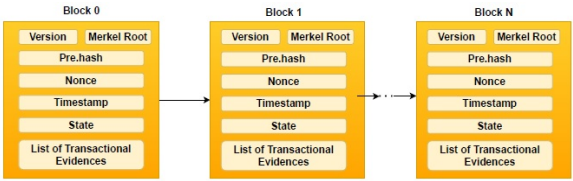

Figure 2: Blockchain

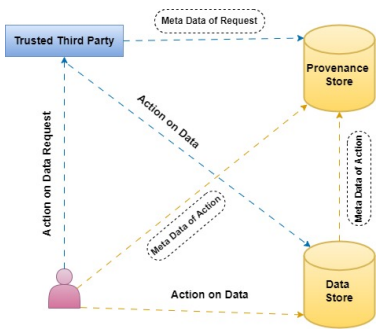

Figure 3: Provenance Capture

## 2.3 Cross Chain

The design of different blockchain systems is heavily influenced by diverse application requirements, leading to challenges in their interoperability and creating isolated data segments that complicate connecting individual systems [30]. For instance, institutions deploying separate blockchains may face challenges when their users need to interact across these systems, requiring a cross-chain interaction model [60]. To address these issues, cross-chain functionality emerges as a solution, allowing for seamless data and asset transfer across different blockchain networks [90]. Cross-chain systems often rely on notary schemes, hash-locking techniques, atomic swaps, side chains, or relay chains to facilitate these interactions [16, 25, 29, 35, 48, 71, 81, 86]. Notary schemes use intermediaries to facilitate transactions between chains, while hash-locking contracts streamline asset exchanges. Atomic cross-chain swaps facilitate asset trading between separate blockchains and ensure that all linked transactions are either fully completed or entirely aborted. Side chains run parallel to main chains, enhancing performance, and relay chains focus solely on data transfer between different chains [14, 26, 32, 54, 85].

## 3 RQ1: Provenance for Single Entity

In this RQ, we consider individuals who are not organization-based and owners of their data. The choice of this domain is motivated by the need to understand how a single entity editing and storing data impacts requirements. An illustrative example of such domains includes cloud storage. The increasing reliance on cloud infrastructure for data storage, sharing, and processing has made data provenance in cloud storage environments a significant research concern [61]. Cloud-based provenance systems are vulnerable to accidental corruption or intentional forgery, prompting the widespread use of blockchains due to their ability to provide secure, transparent, and decentralized data management. For example, ProvChain [47] is a data provenance architecture built on a blockchain that addresses the need for secure data provenance in cloud storage applications. It audits data operations for cloud storage and provides real-time cloud data provenance by monitoring user operations. However, it lacks clarity on the identity and trustworthiness of nodes and auditors on the public blockchain, as well as the consensus mechanism employed. Another method [56] introduces a data provenance architecture using blockchain technology to safeguard data activities in a cloud storage application. It records operation histories as provenance data using OpenStack-based Swift storage, storing data hashes as blockchain transactions. This approach enhances data provenance, contributing to the integrity and security of recorded data activities. Additionally, the study develops Blockchain as a Service (BaaS) for provisioning the Ethereum blockchain platform on an OpenStack private cloud and creates a Dropbox-like storage application using OpenStack Swift for logging user metadata. BlockCloud [75] is an architecture for information provenance enabled by blockchain, designed for the cloud computing framework. It implements a PoS consensus mechanism to decrease computational requirements compared to traditional PoW consensus. The method in [33] proposes a secure data provenance system for cloud storage using Blockchain and InterPlanetary File System (IPFS), ensuring data integrity and privacy, and enhanced availability.

### 3.1 Challenges

The literature on individual provenance presents several limitations, including privacy concerns. For example, in cloud service providers, there's a potential for correlating a specific provenance entry to the data owner, depending on the system's implementation. The presented solutions in the papers are specific and the effectiveness with other cloud storage solutions requires further investigation. Some use public blockchains and the impact of blockchain mining costs on cloud service providers is not explicitly addressed. Moreover, public blockchains are transparent, but the system lacks clarification regarding the identity and rationale behind trusting the nodes and auditor on the public blockchain, as well as the consensus mechanism employed. Few frameworks are in an abstract model state, lacking both access control and searching functionalities. A thorough analysis of existing frameworks for provenance highlights an absence of a unified solution that can thoroughly capture, extract, and query provenance data across these systems. This gap hinders the ability to track the origins, dependencies, and transformations of data, which are crucial for maintaining data integrity, trustworthiness, privacy, and accountability. It is crucial for these frameworks to prioritize privacy-preserving methods for provenance capture and querying functionalities.

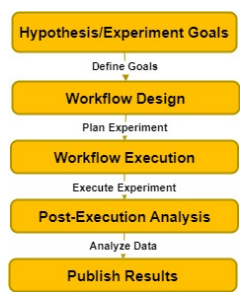

Figure 4: Life cycle of Scientific Workflows

## 4 RQ2: Provenance of Intra-Chain Collaboration

After addressing individual provenance concerns, the focus shifts to domains that necessitate collaboration and the aggregation of information from multiple users. This is especially pertinent in fields where provenance plays a critical role, including scientific collaboration, supply chain management, healthcare systems, machine learning, and digital forensics. These selected domains rather than other areas, are pivotal due to their extensive need for secure data integrity and effective provenance tracking in collaborative and multi-organizational contexts. Additionally, there is a wealth of existing research specifically addressing provenance issues within these domains, providing a robust foundation for further exploration. With this RQ, we analyze these domains, their requirements, and challenges by examining each domain individually. Summary of design considerations can be seen in Table 2.

### 4.1 Scientific Collaboration

Provenance is crucial in scientific research collaborations, but current systems have limitations in data security and collaboration. There's a growing need for tamper-proof storage of scientific data provenance to facilitate collaboration and data sharing. Funding agencies are increasingly requiring researchers to share data, leading to a demand for more open research practices. An example of the life cycle of scientific workflows [50] is illustrated in Figure 4. While existing blockchain solutions provide tamper-proof storage, they lack support for data sharing and complex workflows. Several studies propose solutions customized for managing scientific workflow provenance. For example, BlockFlow [22] uses integrated event listeners, akin to Lineage Chain [65], but constructs the blockchain atop the E-science ecosystem to detect data alterations. SmartProvenance [63] employs threshold-based voting systems and custom smart contracts to authenticate provenance records using the Open Provenance Model. SciBlock [28] introduces a timestamp-based invalidation mechanism supporting workflow modifications. Bloxberg [80] introduces a unique provenance model encompassing configuration details, code, and other data specific to scientific software systems. SciLedger [36] addresses these challenges by offering a blockchain platform for collecting and storing scientific workflow provenance. It supports multiple workflows, complex operations, and has an invalidation mechanism. Earth observation (EO) uses remote sensing to gather data about the Earth's surface and atmosphere. The growing volume of EO data, reaching the Petabyte scale annually, poses challenges for storage and processing in centralized environments. Blockchain's distributed nature is considered a potential solution to these challenges. In [87] The blockchain based EO data management system involves three main components: users, data centers, and the blockchain. Users upload EO datasets to data centers, which utilize a consortium blockchain with Raft and PBFT consensus algorithms to achieve high throughput, low latency, and efficient querying. Data centers store EO data off-chain, while essential information is stored on-chain and managed by smart contracts. Transactions within the blockchain form a Directed Acyclic Graph, enabling efficient traceability, enhancing scalability and interoperability.

### 4.2 Supply Chain Management

Research on blockchain for supply chain provenance is categorized into various sectors, including the pharmaceutical industry, which is pivotal in the global supply chain. This industry oversees the creation, manufacturing, and distribution of vital medications and healthcare products. pharmaceutical supply chain involves stakeholders such as producers, distributors, pharmacies, and healthcare providers. Effective management is crucial to address challenges related to coordination, communication, procurement, storage, shipping, and regulatory compliance. Several key issues impact the pharmaceutical supply chain, including counterfeiting, product tampering, cold chain management, quality control and assurance, demand forecasting, and supply chain resilience [53]. Counterfeit drugs can endanger patient safety by mimicking genuine medications, while maintaining the cold chain is crucial for preserving the efficacy of biology and vaccines. Rigorous quality control processes are essential to ensure the safety and effectiveness of medicines. Accurate demand forecasting helps avoid stockouts or overstock situations, given the limited shelf life of pharmaceuticals. Additionally, developing resilient strategies, such as contingency planning and risk mitigation, is vital to maintaining a continuous supply of medicines despite disruptions from natural disasters, geopolitical events, or other factors.

Another significant sector discussed in the literature is the use of blockchain for authenticating electronic devices within the supply chain, as highlighted by Cui *et al.* [23]. This approach offers a unique device ID for tracking parts to address issues such as counterfeit ICs and cloning. It enables precise tracking of a device's origin and its progression through the supply chain, ensuring its authenticity. The proposed system introduces a confirmation-based ownership transfer mechanism to prevent theft, human errors, and fraudulent entities.

Table 2: Considerations in Blockchain Collaborative Applications for Provenance Across Domains

| Scientific Collaboration | Digital Forensics | Machine Learning | Supply Chain | Healthcare Systems |
| --- | --- | --- | --- | --- |
| Intellectual property | Coordination of investigation stages | Monitoring data gathering for training | Device ownership transfer | Determining data ownership |
| Managing data workflow, private data inputs | Handling multi-modal data | Addressing non-IID data | Illegitimate product registration | Manager of access |
| Flexibility for re-execution | Utilizing AI/ML techniques | Documenting all steps of training | Incentives to share provenance | HIPPA |
| Invalidating tasks | Analyzing encrypted data | Managing statistical heterogeneity | Focus on specific industries | Goals of collaborations |

Hyperledger Fabric is employed for a private blockchain in the prototype, aiming to reduce transaction costs and enhance efficiency. Xu et al. [83] provided a detailed solution and overview of how blockchain can enhance and secure the integrity of the electronic supply chain. However, their solutions lack detailed traceability and ownership information for devices. On the other hand, Islam et al. [38] suggested a method that incorporates Physically Unclonable Functions (PUF) and blockchain to improve the authenticity and traceability of parts in the supply chain. Nonetheless, the transfer of device ownership in their approach is solely initiated and managed by device owners.

Other domains within the supply chain also utilize blockchains to provide platforms for all stakeholders in the food supply chain, from farmers to retailers, allowing them to input and access data, thus ensuring transparency and real-time visibility. This access empowers consumers to obtain reliable information about the food's origin, quality, and safety. Moreover, blockchain's inherent security measures make it a robust tool for detecting and preventing food fraud, as any attempt to alter the recorded data is quickly identified. The methodology outlined in Kumar et al. [42] involves three main modules: Source Tracking, which uses IoT sensors and RFID tags with blockchain to monitor food products from origin to consumption; Quality and Safety Monitoring, which ensures food items' quality and safety by tracking parameters like temperature and humidity, with blockchain for easy tracking and alerts for deviations; and Certification and Compliance, which maintains certification documents on the blockchain for easy verification, thereby enhancing food safety and transparency. PrivChain [52] is proposed as a general framework for supply chain participants to secure sensitive data on the blockchain using zero-knowledge proofs, ensuring provenance and traceability without revealing information to end-consumers or supply chain entities. It allows data owners to provide proofs instead of data and gives incentive to entities to supply valid proofs using Zero Knowledge Range Proofs (ZKRPs) without disclosing exact locations. Offline computation of proofs reduces blockchain overhead, while proof verification and incentive payments are automated through blockchain transactions, smart contracts, and events. LedgerView [66] introduced a system that adds access control views to Hyperledger Fabric, supporting both revocable and irrevocable views with role-based access control. However, it lacks some privacy demands such as anonymity.

### 4.3 Healthcare Systems

Provenance is a key factor in healthcare systems, where in this domain it is the lifecycle of the electronic health record (EHR). Singh et al. [69] developed a blockchain-enabled electronic health record (EHR) healthcare framework utilizing smart contracts to manage a range of stakeholders, including doctors, patients, pathology labs, chemists, and insurance providers. This framework guarantees data privacy, availability, immutability, and authentication while also offering query capabilities. Their findings suggest that blockchain-enabled EHR systems could greatly enhance next-generation healthcare. Abouyoussef et al. [3] proposed platform addresses the critical need for an online-automated system during pandemics, enabling remote collection of symptoms, accurate diagnostics, and secure data sharing in the healthcare system. It utilizes a custom-designed blockchain platform that ensures privacy through group signature and random numbers, supporting anonymity and data unlinkability. A deep neural network based detector, implemented as a smart contract, enables automatic diagnostics with high accuracy. The platform also allows healthcare entities access to symptom and diagnosis data through a consortium-based blockchain architecture. The authors of [27] proposed a blockchain-based clinical data sharing system where local clinics receive medical summaries from regional hospitals and other healthcare centers' electronic medical records. After analyzing the data, local hospitals bundle the health information into blocks and send them to agreement hubs. The hospitals, acting as both validators and endorsers, are responsible for verifying, validating, and endorsing these blocks. Each hospital has the option to either store its patient data on its own ledger or submit it to the blockchain. The system uses customized access control protocols and symmetric cryptography for security but the blockchain configurations are not addressed. The work in [59] introduced a secure method for sharing EHRs via a private blockchain. This method enables multi-user search capabilities and utilizes ciphertext-based attribute encryption to maintain data confidentiality. It ensures detailed access control and prevents unauthorized doctors from uploading false information by enforcing authenticated access. Abdelgalil et al. [1] presents HealthBlock, a framework for securely sharing EHRs while maintaining privacy. It leverages IPFS for distributed storage, Hyperledger Indy for patient control, and Hyperledger Fabric for managing access. HealthBlock enhances current EHR systems by ensuring record integrity and confidentiality, granting patients control over access, supporting anonymous healthcare services, enabling remote attribute verification for telemedicine, and addressing emergency access needs.

### 4.4 Machine Learning

The traceability of machine learning (ML) and deep learning (DL) algorithms is especially vital for establishing dependable and trustworthy AI analytics practices in manufacturing settings. This traceability significantly enhances cybersecurity by identifying and mitigating threats like poisoning attacks against AI systems [70]. Moreover, it facilitates the adoption of adaptable cyber-defense measures by verifying the reliability of the training data used for AI/ML algorithms. The method described in the study [39] is implementing a blockchain-based provenance approach to enhance the security and integrity of data used in machine learning algorithms, specifically for predicting diabetes in this case. The method ensures that the data remains tamper-proof by utilizing blockchain technology to record the history of the data's origin and any subsequent changes. By distributing the data across on-chain and off-chain transactions, the method adds an additional layer of security, as the data is validated and compared between these transactions Proposed solution for tracking AI assets on a public blockchain by Luthi et al. [51] extends the current provenance model to distinguish between datasets, operations, and models, aiming to address privacy, audibility, and fair compensation concerns within federated learning. This updated model is designed to accommodate interacting AI value chains and to monitor datasets and models without necessitating corresponding operations, thus enabling participants to define their operations. The assets are classified into operations, datasets, and models, and their relationships are depicted in a directed acyclic graph (DAG). To maintain confidentiality, encryption is employed, and provenance graphs are constructed to monitor asset usage and facilitate equitable remuneration. Another example is combining Federated Learning (FL) with blockchain which addresses the vulnerabilities of traditional FL frameworks, such as the risk of a single point of failure, while enhancing security and reducing the risk of malicious activities. This integration also boosts worker motivation. FL requires distributed and secure methods for coordinating collaboration among participants, and blockchain has proven effective in managing these aspects [84]. By decentralizing FL applications and replacing the central server with smart contract execution, blockchain further enhances security and mitigates the risk of malicious activities [40, 46]. BlockDFL [62] is a blockchain-based decentralized P2P framework for FL. It employs a voting mechanism and gradient compression to coordinate FL among participants without mutual trust, defending against poisoning attacks. Yang et al. [84] propose a FL framework that evaluates node reputation to combat model poisoning and free-riding attacks. It uses reputation based methods to categorize nodes, reduce communication, and provide incentive for good behavior. This approach outperforms baselines with Non-IID data and remains stable under 50% attacks.

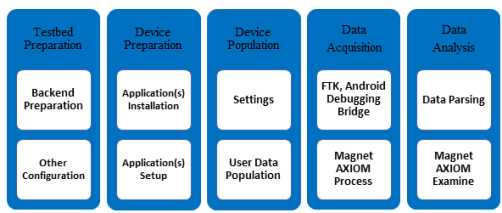

Figure 5: Digital Forensics

### 4.5 Digital Forensics

Digital forensics entails the discovery and interpretation of electronic data intended for legal purposes. It adheres to a structured five-stage methodology as illustrated in Figure 5. Initially, investigators identify sources of evidence and relevant individuals. Subsequently, they preserve Electronically Stored Information (ESI) to prevent any data alteration. The next step involves collecting digital data and creating exact duplicates for detailed analysis. Finally, the findings are compiled into a comprehensive report. This meticulous process ensures the integrity and legal admissibility of evidence, following standards established by organizations. In the context of digital forensics, data provenance plays a crucial role in verifying the legitimacy and origin of data, aiding in identification and reuse, and preserving the integrity of systems [2, 61]. Integrating provenance with digital forensics introduces additional challenges, particularly in terms of security. Digital forensics relies on the integrity and authenticity of evidence, which can be compromised if provenance data is not properly managed. Issues such as tampering, data loss, and unauthorized access can undermine the forensic process and lead to unreliable outcomes. The proposed blockchain-based IoT forensic framework (IoTFC) [45] addresses the need for a comprehensive digital forensics framework. While the strengths of IoTFC include efficient data acquisition and secure verification mechanisms Ahmed et al. [8] propose a cost-effective private blockchain and IPFS system based on Hyperledger for tracking media files as evidence. Their system includes access control, which grants full capabilities to the owner but lacks detailed specifications. On the other hand, Lone et al. [49] and Tsai et al. [78] utilized Ethereum, a public blockchain, to enhance criminal investigations and establish a chain of custody mechanism. In [13], a model uses blockchain for secure evidence storage. Data is organized into blocks, encrypted, and linked. A cover file is created from the previous block's data and encrypted to form a cipher file. Evidence is preprocessed, divided into chunks, and encrypted. These encrypted chunks are embedded into the cipher file to create a steganography file, which is then stored in the blockchain through mining, ensuring integrity and confidentiality. ForensiBlock [12] is a digital forensics solution that tracks all investigation data, including communication records, enabling quick evidence extraction and verification while safeguarding sensitive information. It features new methods of access control, supporting investigation stage changes, and employs a distributed Merkle tree for case integrity verification.

### 4.6 Challenges

Each domain presents its own unique set of obstacles and requirements that need to be carefully addressed to ensure the successful implementation and utilization of blockchain technology. Integrating blockchain into scientific workflow management presents challenges in maintaining intellectual property and confidentiality, ensuring auditing support, and enabling flexibility for re-execution. Additionally, interoperability of systems and data models, as well as managing the complexity of scientific data, require further research and development. These challenges underscore the need for efficient access control, privacy preserving techniques, and scalable solutions to fully leverage blockchain technology in scientific workflows [34]. Nevertheless, the designs of these systems often overlook security measures and the complexity of workflow management, especially with branching, merging, and invalidating tasks, which can become significant challenges. Some systems operate on a public blockchain, which can create privacy problems, and they may struggle with managing private data inputs, lacking robust support for such data. Blockchain designs for supply chains face several challenges, including device ownership transfer, where tracking the movement of devices through various stages of the supply chain must be accurately recorded. Additionally, preventing illegitimate product registration is crucial to ensure that only genuine products are tracked and traced on the blockchain. Another significant challenge is providing incentives for participants to compute and share privacy-preserved provenance data. Implementing healthcare systems can pose several challenges, including establishing data ownership, maintaining patient centricity, controlling access, and addressing privacy concerns. For instance, in diagnostic scenarios, ensuring complete privacy preservation is paramount. It's crucial to prioritize anonymity and data unlinkability to safeguard patient information. Moreover, healthcare data is governed by strict regulations like HIPAA in the United States. Complying with these regulations can be intricate and demands meticulous attention. Additionally, user adoption and practicality remain significant hurdles to overcome. In blockchain-based provenance for ML systems, there is a need for monitoring the gathering of training data and following up on all steps of training to ensure data integrity, transparency, accountability, help identify and correct issues. ML collaborative approaches can be vulnerable to background knowledge attacks, collusion attacks, inference attacks, poisoning attacks, and privacy breaches [64, 67, 74]. Additionally, challenges arise from the data itself, especially when handling different data modalities, statistical heterogeneity, and the non-IID (non-Independent and Identically Distributed) nature of the data. The existing designs on blockchain-based digital forensics often falls short in several key areas, including access control, consideration of attack scenarios, and communication coordination. Effective collaboration in digital forensics requires coordinated efforts across different stages of an investigation, while also addressing the unique needs of each stage. Since blockchains are transparent, any evidence stored on them must be encrypted to protect sensitive information. However, analyzing encrypted data poses a challenge, as it must be done without compromising its integrity or security. Digital forensics frequently involves analyzing data from various sources, such as text, images, and videos, often utilizing AI/ML techniques. Integrating these diverse data modalities into blockchain systems demands specialized techniques and tools to ensure the integrity and accuracy of analysis results. Additional challenges in blockchain digital forensics include the tokenization of artifacts from digital evidence, efficient management of data volume in the chain of custody, parsing forensic sound procedures in blockchain systems, enabling understandable forensic outcomes and reports, ensuring interoperability and addressing cross-border jurisdictions, and establishing a clear timeline of events and chronology [24].

## 5 RQ3: Provenance of Multi-chain Collaboration

Discussing the needs for collaboration and provenance from the previous RQ has brought us to another scenario: one where multiple organizations aim to collaborate, each employing its own blockchain. In the realm of blockchain communication and interoperability, it becomes vital to establish a communication structure between these disparate blockchains. This raises questions about how blockchains interpret each other's data, the involvement of nodes in the communication process, and whether this impacts the consensus mechanism. Interoperability also prompts concerns about data consistency, building trust, and the implications of cross-chain transactions on the security and performance of individual chains. Hwang *et al.*[37] introduced a two-layer organization method for blockchain systems, involving a main blockchain and a side blockchain with the same architecture. This approach allows for effective data sharing within a homogeneous side blockchain, enabling easy data management and reducing redundancy. However, it struggles with expansion to heterogeneous participant blockchains, where different data structures prevent direct communication and interaction between systems, making secure data sharing in a heterogeneous environment challenging. Furthermore, The ARC [88] system presents an innovative approach to cross-chain communication for blockchain communication and interoperability. While ARC focuses on security and provides a clear system description, it lacks a thorough evaluation and detailed implementation discussion. Improvements could include a detailed evaluation, better implementation discussions, and consideration of alternative trust models for participants. To address the challenges of achieving unified verification mechanisms for shared data and protecting the privacy of sensitive data owners without permission control, SynergyChain [21] introduces a three-tier architecture based on blockchain. Its aim is to enable data sharing and resolve data access controllability in a multichain environment. SynergyChain has demonstrated its ability to support data sharing reliably and efficiently, reducing data query latency compared to sequentially requesting multichain data. Current solutions for data sharing among multiple organizations often overlook the protection of sensitive data. In contrast, SynergyChain aggregates data in a multichain system to facilitate data sharing among multiple institutions while addressing these issues. Regarding cross-chain provenance, the Vassago [31] system stands out for its efficient and authenticated provenance queries. Despite its strengths in efficient query mechanisms and high throughput, Vassago encounters challenges such as resource inefficiency for repeated queries and limited fault tolerance. These challenges include fairness and freshness concerns, likability and indistinguishability issues, trust concerns, and the process of conducting independent queries. Enhancements for Vassago could involve implementing a Trusted Execution Environment (TEE) for query authenticity and enabling modifications of smart contracts based on consensus among all participating nodes in the Dependency Blockchain (DB). ForensiCross [11], the first cross-chain solution for digital forensics, uses BridgeChain to facilitate interactions between private blockchains via a novel communication protocol. It ensures logging, access control, provenance extraction, and synchronization of investigative stages. Nodes validate transactions across blockchains, requiring unanimous agreement for progression. The system manages investigation stages, assigns access privileges, and secures data retrieval and uploading. Provenance is verified through a novel Merkle tree construction, enhancing security.

### 5.1 Challenges

The current literature on the provenance of multichain and cross-chain scenarios is limited. Addressing this gap involves understanding the technical complexities and collaborative challenges inherent in these contexts. One major challenge is preventing attacks that aim to reorganize a blockchain's transaction history, especially in collaborative settings with multiple blockchains. This risk is heightened due to differing consensus mechanisms or validation rules across blockchains. Ensuring fairness is crucial, particularly when different organizations are simultaneously transacting across multiple blockchains. Fairness in this context means providing equal opportunities for all organizations to participate in cross-chain transactions. Designing protocols that uphold fairness is essential to prevent biases in collaborative settings. Privacy leakage is another significant concern. It occurs when sensitive information is inadvertently exposed during cross-chain transactions. Mitigating this risk requires robust encryption and access control mechanisms. Efficient communication between different blockchains is vital for seamless collaboration. Optimizing cross-chain transaction performance minimizes latency and resource consumption. This optimization involves designing efficient communication protocols and configuring network settings. Trust is fundamental in cross-chain communication, requiring trust between blockchain networks with varying security levels. Building trust involves verifying data authenticity, ensuring communication channel reliability, and trusting consensus mechanisms. Ensuring fairness also involves providing equal opportunities for all organizations to access and interact with the blockchain network. Freshness concerns relate to timely data processing to maintain information integrity. Scalability is crucial, requiring determining the number of nodes and criteria for selecting them to ensure scalability under heavy loads. Capturing provenance is challenging, and conducting independent queries without compromising information integrity or security is essential. Implementing secure and efficient query mechanisms is necessary, especially for large-scale datasets.

## 6 Research Directions

In this section, we outline key research directions for advancing blockchain-based provenance systems.

### 6.1 Design Considerations

The designs of blockchain for provenance vary, prompting considerations for their implementation. The following factors can serve as stepping stones for enhancing future designs:

***Blockchain Choice:*** The decision regarding the blockchain type, whether private or public, holds significant importance. Factors such as participant limitations and privacy considerations often steer towards private blockchains. Furthermore, the choice between crafting a new blockchain or utilizing existing ones presents its own set of challenges. Developing a blockchain involves careful consideration of consensus mechanisms, node incentives, and participant categorization, whether they are individuals or organizations. While creating a new blockchain tailored to specific needs may seem beneficial, it can introduce compatibility issues with existing tools, necessitating the redesign of all related components.

***Domain:*** The specific domain of application plays a crucial role in shaping the design considerations for provenance blockchains, as demonstrated in section 4. Regardless of the domain, several key aspects must be addressed. These include synchronizing processes with collaborators, defining clear roles, considering various phases of the domain, tailoring solutions to meet domain-specific needs, determining whether analysis should be on or off the blockchain, supporting diverse data types, ensuring regulatory compliance, addressing privacy concerns, mitigating different types of attacks, integrating with various tools, and meeting processing and security requirements. Each of these considerations is essential for developing effective and relevant blockchain solutions for provenance tracking.

***Access Control:*** It's crucial to define the entities participating in the blockchain and determine their access levels. This can involve employing traditional access control methods like attribute-based access control (ABAC) or role-based access control (RBAC), or even more sophisticated models such as deep learning [10]. However, access control strategies may need to be customized to suit the specific requirements of the domain. Additionally, in collaborative environments, there arises the question of who should be responsible for implementing access control and how consensus on these measures is reached.

***Provenance Capture:*** Capturing provenance is crucial, demanding careful attention to communication details, data significance, and storage decisions. While some approaches exclusively capture provenance on the blockchain, others store this information externally. It's important to define provenance within the specific domain and determine its purpose to make informed capture choices early in the process.

***Provenance Query:*** Whenever provenance is captured, it serves a purpose, necessitating its retrieval. Querying methods can vary; sometimes, precise information is extracted, while other times, a batch of information containing the required data is retrieved. It's crucial to clarify the meaning of a query and define our methodology for it. Additionally, having an alternative validation method for queries can prove beneficial in certain scenarios.

***Evaluation:*** The evaluation of proposed methods in depends on the specific method and its impact on system overheads, including both time and storage. Various aspects can be evaluated based on the design, such as throughput, retrieval latency of provenance, storage performance overhead, query service interaction with the provenance database, overhead for provenance data upload, and validation time. Additionally, considerations for designing a new blockchain or making changes to the structure of an existing blockchain, such as consensus algorithms, hardware impact, block formation, transaction processing time, difficulty level, load, and network size, are also important factors to evaluate.

### 6.2 Future Work

Building upon the challenges outlined in preceding sections, that need to be addressed, we aim to underscore several under-explored or inadequately investigated topics for future research. One area is handling repeated queries, where identical queries are frequently made, leading to redundant data retrievals. This redundancy can cause inefficiencies and increased latency, making it crucial to develop methods for handling such requests while preserving privacy. This is particularly important in blockchain provenance to ensure timely and secure retrieval of provenance information, thereby maintaining data integrity and trustworthiness. Optimizing repeated query handling can reduce unnecessary data transfers, lower network load, and provide quicker access to provenance information without compromising privacy. Another important topic is managing multi-modal data, which includes various types such as text, images, and videos. Different data types require unique tokenization and methods to ensure their uniqueness, essential for accurate provenance tracking. Using multi-modal machine learning methods for analysis can improve insights across various domains. Effectively managing multi-modal data allows the provenance system to robustly handle diverse inputs, maintain data integrity and uniqueness, and leverage advanced analytical techniques to provide comprehensive and reliable provenance information. This approach enables more detailed tracking and verification, enhancing the overall reliability and effectiveness of blockchain-based provenance systems. Future directions for multi-chain and cross-chain provenance highlight the need for a unified solution that encompasses communication methods, provenance capture, and query mechanisms. Current approaches are insufficient, and there is a critical need for methods that also prioritize security and privacy. Finally, there is a need for multi-chain methods tailored to specific domains or for handling cross-domain collaborations while addressing domain-specific requirements. Each domain has unique challenges and requirements that must be met to ensure effective and secure collaboration. Tailored solutions can enhance the accuracy, efficiency, and security of provenance tracking in various applications, ensuring that the specific needs of each domain are adequately addressed.

## 7 Conclusions

This paper aims to provide a comprehensive and fresh perspective on blockchains designed specifically for provenance, exploring the distinct requirements and challenges associated with blockchain provenance across various contexts, including individual applications, single-chain systems, and multi-chain environments. To achieve this, the paper breaks down these needs into three key RQs, each addressing a different aspect of blockchain provenance. It reviews existing work on each research question, offering a thorough analysis of the literature and current practices, highlighting drawbacks and limitations, and providing a critical assessment of areas where improvements are needed. Furthermore, the paper explores future directions for research and development in blockchain-based provenance, suggesting new areas of investigation.

## 8 ACKNOWLEDGMENTS

This work was funded by NSF grant CCF-2131509